\documentclass[secnumarabic,amssymb,amsmath,nobibnotes,aps] {revtex4-1}
\usepackage[all, knot]{xy}
\usepackage[T1]{fontenc}
\usepackage[latin9]{inputenc}

\newtheorem{theorem}{Theorem}

\newtheorem{lemma}{Lemma}
\newtheorem{corollary}{Corollary}

\usepackage[all, knot]{}
\begin{document}

\title{Towards superconformal and quasi-modular representation of exotic smooth $\mathbb{R}^4$ from superstring theory I}

\author{Torsten Asselmeyer-Maluga}

\email[REV\TeX{} Support: ]{torsten.asselmeyer-maluga@dlr.de}

\affiliation{German Aero Space Center, Rutherfordstr. 2, 12489 Berlin}

\author{Jerzy Kr\'ol}

\email[REV\TeX{} Support: ]{ iriking@wp.pl}

\affiliation{University of Silesia, Institute of Physics, ul. Uniwesytecka 4,
40-007 Katowice}

\begin{abstract}
We show that superconformal ${\cal N}=4,2$ algebras are well-suited to represent some invariant constructions characterizing exotic $\mathbb{R}^4$ relative to a given radial family. We examine the case of ${\cal N}=4, \hat{c}=4$ (at $r=1$ level) superconformal algebra which is realized on flat $\mathbb{R}^4$ and curved $S^3\times \mathbb{R}$. While the first realization corresponds naturally to standard smooth $\mathbb{R}^4$ the second describes the algebraic end of some small exotic smooth $\mathbb{R}^4$'s from the radial family of DeMichelis-Freedman and represents the linear dilaton background $SU(2)_k\times \mathbb{R}_Q$ of superstring theory. Via $\sigma$-model realization of linear dilaton one derives various results driven by gravity of exotic $\mathbb{R}^4$ in the regime of closed string theory. 

From the modular properties of the characters of the algebras one derives Witten-Reshetikhin-Turaev and Chern-Simons invariants of homology 3-spheres. These invariants are represented rather by false, quasi-modular, Ramanujan mock-type functions. Given the homology 3-spheres one determines exotic smooth structures of Freedman on $S^3\times \mathbb{R}$. In this way the fake ends are related to the SCA ${\cal N}=4$ characters.

The case of the ends of small exotic $\mathbb{R}^4$'s is more complicated. One estimates the complexity of exotic $\mathbb{R}^4$ by the minimal complexity of some separating from the infinity 3-dimensional submanifold. These separating manifolds can be chosen, in some exotic $\mathbb{R}^4$'s, to be homology 3-spheres. The invariants of such homology 3-spheres are, again, obtained from the characters of SCA, ${\cal N}=4$. 

Next we take into account the modification of the algebra of modular forms due to the noncommutativity of the codimension-one foliations of the homology 3-spheres. These foliations are cobordant to the codimension-one foliations of the 3-sphere. The cobordism class of the foliation represents the link between the algebraic end and the topological end of exotic $\mathbb{R}^4_k$ in the given radial family. Then, the modification of modular forms is represented by the Connes-Moscovici construction where the Godbillon-Vey class acts via the cyclic cohomology class of certain Hopf algebra. The action deforms the modular forms towards the quasi-modular mock theta functions, and is understood as the action of the noncommutative geometry of the foliation.

\end{abstract}


\maketitle

\section{Introduction}
Exotic smooth $\mathbb{R}^4$'s are quite strange objects. On the one hand they are Riemannian smooth 4-manifolds, which are all hemeomorphic to the topological $\mathbb{R}^4$ being non-diffeomorphic to it. On the other hand they place themselves at the special position where various constructions from lower and higher dimensional topology, differential and algebraic geometry behave critically. This is particularly remarkable that one can interpret and relate these smooth structures into constructions from quantum field theory, quantum physics or noncommutative geometry. The subject is in no way understood completely rather it is just scratched, especially if one looks for proper 4-exotic smooth invariants as derived from physical theories, like a kind of topological quantum field theory, and which would be involved essentially in physical contexts. There were, however, proposed some relative results in this direction. One can briefly summarize them as: \emph{given a fixed radial family of continuum many small exotic $\mathbb{R}^4$'s one can interpret the family in the algebraic context of $\sigma$-models, Kac-Moody algebras, gerbes and exact backgrounds of superstring theory such that, the change between the members of the family is expressed as the corresponding change of the elements of the above structures.} Moreover, also direct, path integral calculations of the QG effects were presented recently \cite{Asselmeyer-Maluga2010,Asselm-Krol-2011}. 
Since the first proofs of existence of exotic $\mathbb{R}^4$ appeared people were searching for the explicit coordinate-like presentations of these open smooth 4-manifolds. The Bi{\u z}aca-Gompf topological constructions generate families of local patches on exotic $\mathbb{R}^4$ however still the effective, from the point of view of calculus and applications in physics, presentation is missing. We know that some exotic $\mathbb{R}^4$'s allow, at least formally, for finitely many local charts in their atlases (e.g. the simplest exotic $\mathbb{R}^4$ of Gompf allows formally three local patches). The difficulty with effective description by atlases seems to be fundamental and touches partially unknown mathematics. Global coordinate map cannot exist on any exotic $\mathbb{R}^4$ - otherwise it would be diffeomorphic to the standard smooth. The difficulty with local presentation on the one hand and the existence of finite covers on the other, caused that the proposition was pushed forward namely, that global standard coordinate axes undergo a kind of entanglement, when on exotic $\mathbb{R}^4$ \cite{Krol:04a,Krol:04b,Krol:2005}. The entanglement could be analyzed by algebraic means of some quantum field theory and quantum mechanics. Let us note here that in the Witten approach to compact exotic 4-manifolds, Donaldson's polynomials were calculated as correlation functions of some (twisted) 4-dimensional Yang-Mills supersymmetric theory \cite{Wit:89}. Probably, supersymmetry is the indication what tools should be used in searching for the proper description of fake open 4-manifolds (see part II of that work).  

In this paper we propose and analyze suitable environment which partially validates such thinking. The tools derive from 2d CFT and string theory. Owing the fact that superstring theory represents very rich mathematics, some activity towards relating the theory with small exotic $\mathbb{R}^4$'s has taken place recently. In this way one obtains new 4-dimensional results from superstrings which are quite independent on the existing techniques like compactifications or the like. But also conversely, superstring theory supplies mathematics which is suitable for analyzing open exotic 4-smoothness \cite{AssKrol2010ICM,AsselmKrol2011f}.  

First, we describe known realization of some superconformal theories in standard $\mathbb{R}^4$ such that the superconformal fields serve as the coordinates. These are free fields of ${\cal N}=4, \hat{c}=4$ algebra represented in flat (standard) $\mathbb{R}^4$. The vertex operators of the bosonic fields describe four geometric axes of the standard $\mathbb{R}^4$. Next, we look for an analogous superconformal algebraic setting which might be suitable for the case of (some) small exotic $\mathbb{R}^4$'s. When turning to the codimension-one foliations of 3-sphere with non-vanishing Godbillon-Vey invariant, as derived from small exotic $\mathbb{R}^4_k$ from the fixed radial family, one finds the algebraic end $SU(2)_k\times \mathbb{R}$ of $\mathbb{R}^4_k$. Then, the same ${\cal N}=4, \hat{c}=4$ algebra can be represented on this algebraic end. This is realized by the WZW $\sigma$-model in this target and represents exact superstring background of closed superstring theory - the linear dilaton $SU(2)_k\times \mathbb{R}_{\phi}$ background. The appearance of superconformal field theories and algebras rises a more general question like the relevance of this kind of structures in description of exotic smooth $\mathbb{R}^4$'s beyond the algebraic level, and about their possible relation to quantum physics. In this way we have the change of exoticness on $\mathbb{R}^4$ encoded in the level $k$ of the WZW model $SU(2)_k\times \mathbb{R}_Q$. The coordinate axes represented by free fields in the standard smooth case, are shifted to the interacting fields of some quantum CFT in the exotic case.

That is why after brief presentation, following Ref. \cite{AsselmKrol2011f}, how to obtain 4-dimensional physical results based on 'exotic' parts of superstring backgrounds and algebraic exotic ends, we attempt to explain the relation of the algebraic ends with actual topological ends of $\mathbb{R}^4_k$. There appears that the level of modularity of the partition functions of CFT as assigned to exotic $\mathbb{R}^4$'s becomes important. Next, this issue is more carefully studied. We turn again to the superconformal algebras ${\cal N}=4$ via their so called massless and mass characters. The massless, BPS characters are nearly modular in the precise sense of mock Ramanujan quasi-modular functions. The recognition and understanding of this phenomenon was possible due to the progress mainly made by Zweger in his thesis \cite{Zwegers2002PHD}. Consequently, one is able to calculate the Witten-Reshetikhin-Turaev, Chern-Simons and others quantum invariants of various homology 3-spheres (see e.g. Refs. \cite{LawrZag1999,Hikami2005,Hikami2005a,Zagier2006}). On the other hand, the homology 3-spheres are inherently involved in the construction of 'fake' smooth structures of Freedman on $S^3\times \mathbb{R}$ and also in the description of exotic $\mathbb{R}^4$'s. Thus, the relation between the characters of SCA, ${\cal N}=4$ and fake $S^3\times_{\Theta} \mathbb{R}$ is formulated as Theorem \ref{Modular-1}. 

Next, we turn to the deeper level of the codimension-one foliations of $S^3$ and homology 3-spheres and observe that one foliation is cobordant with the other, hence giving a way to shift between algebraic and topological ends for exotic $\mathbb{R}^4$ from the fixed radial family. The foliations in question are non-trivial examples of noncommutative spaces by Connes. The approach of Connes and Moscovici explains how the structure of the algebra of modular forms of all levels, ${\cal M}$, is to be modified due to the GV class of the foliation. First, one builds ${\cal H}_1$ the Hopf algebra whose cyclic cocycle represents the GV class. Next, the ${\cal H}_1$ acts on the crossed product ${\cal M}\ltimes {\rm GL}^+(2,\mathbb{Q})$. This allows for the formulation of Theorem \ref{Modular-2}. Brief presentation of the Connes-Moscovici construction is included in \ref{app}.    

More thorough understanding of the quasi-modular forms as characterizing exotic 4-spaces is presented in the accompanying to this paper. It appears that the Seiberg-Witten Lagrangian acquires calculable quasi-modular corrections due to the exotic $\mathbb{R}^4$ in a special limit.

\section{The radial family of exotic $\mathbb{R}^4$'s and codimension-1 foliations of $S^3$}\label{RF}
Here we review the results of Ref. \cite{AsselmeyerKrol2009}. The standard $\mathbb{R}_{std}^{4}$ is the only smooth differential structure which agrees with the topological product of axes $\mathbb{R}\times\mathbb{R}\times\mathbb{R}\times\mathbb{R}$.
An exotic $\mathbb{R}^{4}$ is the same topological 4-manifold  $\mathbb{R}^{4}-$ but with a different (i.e. non-diffeomorphic) smooth structure. This is possible only for $\mathbb{R}^{4}$ which is the only Euclidean space $\mathbb{R}^{n}$ with an exotic smoothness structure \cite{Asselmeyer2007,Scorpan2005}. We deal exclusively with \emph{small} exotic $\mathbb{R}^4$. These arise as the result of failing $h$-cobordism theorem in 4-d (see, e.g. Ref. \cite{Asselmeyer2007}). The others, so called large exotic $\mathbb{R}^4$, emerge from failing the smooth surgery in 4 dimensions. 
Small exotic $\mathbb{R}^{4}$ is determined by the compact 4-manifold $A$ and attached several \emph{Casson handles} CH's. $A$ is the Akbulut cork and CH is built from many stages towers of immersed 2-disks. These 2-disks cannot be embedded and the intersection points can be placed in general position in 4-d in separated double points. Every CH has infinite many stages of intersecting disks. However, CH is topologically the same as (homeomorphic to) open 2-handle, i.e. $D^2\times \mathbb{R}^2$. If one replaces CH's, from the above description of small exotic $\mathbb{R}^{4}$, by ordinary open 2-handles (with suitable linking numbers in the attaching regions) the resulting object is standard $\mathbb{R}^{4}$. The reason is the existence of infinite (continuum) many diffeomorphism classes of CH, even though all are topologically the same.

The main technical ingredient of the relative approach to small 4-exotics is the relation of these with some structures defined on a 3-sphere. This $S^3$ should be placed as a part of the boundary of the Akbulut cork in $\mathbb{R}^4$. The Akbulut cork has the boundary which is, in general, a closed 3-manifold with the same homologies as ordinary 3-sphere -- the homology 3-sphere. The parameterized by the radii $\rho\in \mathbb{R}$ of $(S^4,\rho)$ a family of exotic $\mathbb{R}^{4}_{\rho}$ exists. Each exotic $\mathbb{R}^{4}_{\rho}$ from the family is the open submanifold of standard $\mathbb{R}^4$. For different $\rho$ taken from the standard Cantor set of reals, corresponding exotic $\mathbb{R}^4_{\rho}$'s are not diffeomorphic. This is what is called the \emph{radial family} of continuum many different non-diffeomorphic small exotic $\mathbb{R}^4$'s, discovered and described by DeMichelis and Freedman \cite{DeMichFreedman1992}. If $CS$ denote the standard Cantor set as a subset of $\mathbb{R}$, then the crucial result is:

\begin{theorem}(Ref. \cite{AsselmeyerKrol2009})\label{th1}
\label{thm:codim-1-foli-radial-fam}Let us consider a radial family $R_{t}$ of small exotic $\mathbb{R}_{t}^{4}$ with radius $\rho$ and $t=1-\frac{1}{\rho}\subset CS \subset[0,1]$ induced from the non-product h-cobordism $W$ between $M$ and $M_{0}$ with Akbulut cork $A\subset M$ and $A\subset M_{0}$, respectively.
Then, the radial family $R_{t}$ determines a family of codimension-one foliations of $\partial A$ with Godbillon-Vey invariant ${\rho}^{2}$. Furthermore, given two exotic spaces $R_{t}$ and $R_{s}$, homeomorphic but non-diffeomorphic to each other (and so $t\not=s$), then the two corresponding codimension-one foliations of $\partial A$ are non-cobordant to each other.
\end{theorem}
$M$ and $M_0$ are compact  non-cobordant 4-manifolds, resulting from the failure of the 4-d $h$-cobordism theorem. Also conversely:
\begin{corollary}(Ref. \cite{AsselmeyerKrol2009}) Given a radial family of small exotic $\mathbb{R}^4$'s then, any class in $H^3(S^3,\mathbb{R})$ induces some small exotic $\mathbb{R}^4$ from the family, where $S^3$ lies at the boundary $\Sigma = \partial A$ of the cork $A$. 
\end{corollary}
For another radial family of exotic $\mathbb{R}^4$'s the same cohomology class determines different nondiffeomorphic exotic $\mathbb{R}^4$. Hence, the foliation is not the diffeomorphism invariant of the exotic $\mathbb{R}^4$.
However, such relativization of 4-exotics to some radial family and to the foliations of $S^3$ and 3-rd real cohomologies of the 3-sphere is the source of variety of further mathematical results and their applications in physics (see, e.g. Refs. \cite{AsselmeyerKrol2009,AsselmeyerKrol2009a,AssKrol2010ICM,AsselmeyerKrol2010,Asselmeyer-Maluga2010,AsselmeyerKrol2011,AsselmeyerKrol2011b,AsselmKrol2011c,AsselmKrol2011d,AsselmKrol2011f,Krol2010,Krol2010b,Krol2011c,Krol2011d}).  

\section{2d CFT and the algebraic end of small Exotic $\mathbb{R}^4$}\label{2dCFT}

In the case of integral \emph{$H^{3}(S^{3},\mathbb{Z})$}
one yields the relation of exotic $\mathbb{R}_{k}^{4}$, $k[\:]\in H^{3}(S^{3},\mathbb{Z})$, $k\in\mathbb{Z}$ with the WZ term of the $k$ WZW model on $SU(2)$.
This is because the integer classes in $H^{3}(S^{3},\mathbb{Z})$
are of special character. Topologically, this case refers to flat
$PSL(2,\mathbb{R})-$bundles over the space $(S^{2}\setminus\left\{ \mbox{\mbox{k} punctures}\right\} )\times S^{1}$
and due to the Heegard decomposition one obtains the relation \cite{AsselmeyerKrol2009}:
\begin{equation}
\frac{1}{(4\pi)^{2}}\langle GV(\mathcal{F}),[S^{3}]\rangle=\frac{1}{(4\pi)^{2}}\,\intop_{S^{3}}GV(\mathcal{F})=\pm(2-k)\label{eq:integer-GV}\end{equation}
the sign depends on the orientation of the fundamental class $[S^{3}]$.
We can interpret the Godbillon-Vey invariant of the foliation of $S^3$ as WZ term.
Namely we consider a smooth map $G:S^{3}\to SU(2)$ and 3-form
$\Omega_{3}=Tr((G^{-1}dG)^{3})$ so that the integral\[
\frac{1}{8\pi^{2}}\intop_{S^{3}=SU(2)}\Omega_{3}=\frac{1}{8\pi^{2}}\intop_{S^{3}}Tr((G^{-1}dG)^{3})\in\mathbb{Z}\]
is the winding number of $G$. Thus indeed every Godbillon-Vey class with integer
value like (\ref{eq:integer-GV}) is generated by a 3-form $\Omega_{3}$.
Therefore the Godbillon-Vey class is the WZ term of the $SU(2)_{k}$ WZW model. The foliation of $S^3$ with this GV class is generated by some exotic $\mathbb{R}^4$, namely $\mathbb{R}^4_k$. Thus, we see that \emph{the structure of exotic $\mathbb{R}_{k}^{4}$'s, $k\in\mathbb{Z}$ from a given radial family determines the WZ term of the $k-2$ WZW model on $SU(2)$.}

This WZ term is required by the cancellation of the quantum anomaly
due to the conformal invariance of the classical $\sigma$-model on
$SU(2)$. Thus we have a way how to obtain this cancellation term
from smooth 4-geometry: when a smoothness of the ambient 4-space,
in which $S^{3}$ is placed as a part of the boundary of the cork,
is precisely the smoothness of exotic $\mathbb{R}_{k}^{4}$, then the WZ term of the classical $\sigma$-model with target $S^{3}=SU(2)$, i.e. $SU(2)_{k}$ WZW,
is generated by this 4-smoothness. The important correlation follows (see, e.g. Ref. \cite{AsselmKrol2011f}):

\emph{The change of smoothness of exotic $\mathbb{R}_{k}^{4}$ to exotic
$\mathbb{R}_{l}^{4}$, $k,\, l\in\mathbb{Z}$ both from the given radial
family, corresponds to the change of the level $k$ of the WZW model
on $SU(2)$, i.e. $k\,{\rm WZW\to}l\,{\rm WZW}$, or $S^3_k \to S^3_l$.}

The end of the exotic $\mathbb{R}_{k}^{4}$ i.e.
$S^{3}\times\mathbb{R}$ cannot be standard smooth and it
is in fact fake smooth $S^{3}\times_{\Theta_{k}}\mathbb{R}$, though the actual smoothness of it is more complicated than this of fake $S^{3}\times_{\Theta}\mathbb{R}$ invented by Freedman \cite{Fre:79}.
So we have determined, via WZ term, the geometry of $SU(2)_{k-2}\times\mathbb{R}$
as corresponding to the exotic geometry of the end of $\mathbb{R}_{k}^{4}$ as the member of the given radial family.
Thus, the change of smoothness on $\mathbb{R}^4$, from standard to exotic as the member of the radial family, corresponds to the change of the geometry of the end, from $S^3\times \mathbb{R}$ to $SU(2)_k\times \mathbb{R}$. This is schematically presented in the Fig. \ref{fig-2} below. 

\begin{figure}[ht]
\centering
\begin{tabular}{cc}
a)  \xymatrix{
 & \mathbb{R}^4_k 
     \ar[ld]^{j_5}
     \ar[rd]_{j_1} 
     \\ S^3_k \times \mathbb{R} \ar[d]_{j_2} && S^3_k \ar[d]_{j_4}
  \\ S^3 \times \mathbb{R}  \ar[rd] \ar[rr]^{Al}
    && S^3_k \times \mathbb{R}  
  \\ & \mathbb{R}^4 \ar[lu]^{\rm{end(\mathbb{R}^4)}} \ar[ru]_{j_3}}

   & \label{diag3} 
   b) \xymatrix{\\ \\ &&
S^3 \times \mathbb{R} \ar[rr]^{Al} && S^3_k \times \mathbb{R}
}  
\end{tabular}     
\caption {The net of correspondences in a fixed radial family: a) $j_1$ generates the $\rm{WZ}_k$-term from exotic $\mathbb{R}^4_k$ via $GV$ invariant of the codim.-1 foliation of $S^3$; $j_5$ is the product by $\mathbb{R}$ of $S^3_k$ assigned to exotic smooth $\mathbb{R}^4_k$; $j_2$ is essentially forgetting the fuzzy 3-sphere structure; $\rm{end(\mathbb{R}^4)}$ assigns the standard end to $\mathbb{R}^4$ and conversely; $j_4$ assigns product by $\mathbb{R}$; $j_3$ shifts $S^3$ to the fuzzy 3-spheres; in this way we have the shift $Al$ between standard end and its 'algebraic' counterpart along with the change of smoothness on $\mathbb{R}^4$, as in b).}\label{fig-2}

\end{figure}
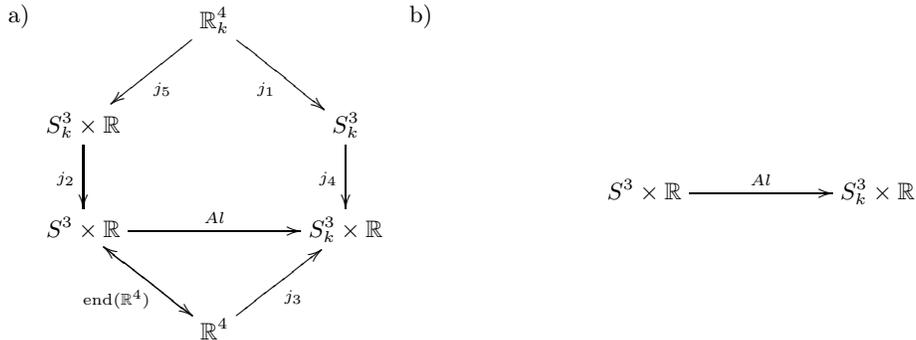

The important thing for us is the replacement as in Fig. \ref{fig-2}b, along with the changes in smooth structures, which allows for superconformal algebraic treatment. The emerging algebraic structure of $SU(2)_k\times \mathbb{R}$ assigned to exotic $\mathbb{R}^4$ we call an \emph{algebraic end} of exotic $\mathbb{R}^4_k$ in the fixed radial family \cite{AsselmKrol2011f}.

\section{${\cal N}=4, \hat{c}=4$ superconformal algebra and its two realizations}\label{4SCFT}
This sections serves as the explanation, on the algebraic level, of the shift $j_3:\mathbb{R}^4 \to S^3_k\times \mathbb{R}$ from the Fig. \ref{fig-2}a, or $Al:S^3 \times \mathbb{R} \to S^3_k \times \mathbb{R}$ from the Fig. \ref{fig-2}b. Throughout this section we understand the algebraic structure of $S^3_k\times \mathbb{R}$ as corresponding to exotic 4-geometry of $\mathbb{R}^4_k$.

Let us, following Ref. \cite{Antoniadis94}, consider the superconformal algebra ${\cal N}=4$ $\hat{c}=4n$ which is defined by the stress energy tensor $T(z)$, the supercurrents $G_a(z), a=1,2,3,4$ and $SU(2)_n$ Kac-Moody currents at the level $n$, i.e. $S_i(z), i=1,2,3$. The following OPE relations emerge by closing the algebra:
 \begin{equation}
\begin{array}{c}
T(z)T(w)\sim \frac{3\hat{c}}{4(z-w)^4}+\frac{2T(w)}{(z-w)^2}+\frac{\partial T(w)}{(z-w)}\\[5pt]
T(z)G_a(w)\sim \frac{3G_a(w)}{2(z-w)^2}+\frac{\partial G_a(w)}{(z-w)} \\[5pt]
T(z)S_i(w)\sim \frac{S_i(w)}{2(z-w)^2}+\frac{\partial S_i(w)}{(z-w)} \\ [5pt]
G_i(z)G_j(w)\sim \delta_{ij}\frac{\hat{c}}{(z-w)^3}-4\epsilon_{ijl}\frac{S_l(w)}{(z-w)^2}+\delta_{ij}\frac{2T(w)}{(z-w)}, i,j,l=1,2,3,4 \\[5pt]
S_i(z)G_j(w)\sim \frac{1}{2(z-w)}(\delta_{ij}G_4(w)+\epsilon_{ijl}G_l(w)),i,j,l=1,2,3,4\\[5pt]
S_i(z)S_j(w)\sim -\delta_{ij}\frac{n}{2(z-w)^2}+\epsilon_{ijl}\frac{S_l(w)}{(z-w)}
\end{array}\label{SU-Con-Alg}
\end{equation}
where $n=1$ for $\hat{c}=4$, i.e. there are three $S_i$ currents of $SU(2)_1$. 

Then we represent this ${\cal N}=4,\hat{c}=4$ super CFT algebra by coordinate bosonic fields $\Phi_a, a=1,2,3,4$ and their $U(1)$ currents, given by the $\sigma$-model with target $\mathbb{R}^4$. One has:
\begin{equation}
\begin{array}{c}
T=-\frac{1}{2}[(\partial H^+)^2+(\partial H^-)^2-PP^{\dag}-\Pi\Pi^{\dag}] \\[5pt]
G=\frac{G_1+iG_2}{\sqrt{2}}=-(\Pi^{\dag}e^{-\frac{i}{\sqrt{2}}H^-}+P^{\dag}e^{\frac{i}{\sqrt{2}}H^-})e^{\frac{i}{\sqrt{2}}H^+} \\[5pt]
\tilde{G}=\frac{G_4+iG_3}{\sqrt{2}}=(\Pi^+e^{\frac{i}{\sqrt{2}}H^-}-Pe^{-\frac{i}{\sqrt{2}}H^-})e^{\frac{i}{\sqrt{2}}H^+} \\[5pt]
S_3=\frac{1}{\sqrt{2}}\partial H^+\:,\;S_{\pm}=e^{\pm i\sqrt{2}H^+}\;.
\end{array}\label{Flat-1}
\end{equation}
Here, \begin{equation*}
\begin{array}{c}
P=J_1+iJ_2,\;\; P^{\dag}=-J_1+iJ_2 \\[5pt]
\Pi=J_4+iJ_3\;\; \Pi^{\dag}=-J_4+iJ_3
\end{array}\label{Flat-2}
\end{equation*}
and $J_a=\partial \Phi_a,\,a=1,2,3,4$ are bosonic $U(1)$-currents, and free fermions are written in terms of two bosons, $H^+,H^-$. Also, the decomposition of the $SO(4)_1$ fermionic currents, $\Psi_i\Psi_j$ in terms of two $SU(2)_1$ currents $S_i,\tilde{S}_k$ was performed, which reads:
\begin{equation*}
\begin{array}{c}
S_i=\frac{1}{2}(\Psi_4\Psi_i+\frac{1}{2}\epsilon_{ijl}\Psi_j\Psi_l)\to (\frac{1}{2}\partial H^+,e^{\pm i\sqrt{2}H^+}) \\[5pt]
\tilde{S}_k=\frac{1}{2}(-\Psi_4\Psi_k+\frac{1}{2}\epsilon_{kjl}\Psi_j\Psi_l)\to (\frac{1}{2}\partial H^-,e^{\pm i\sqrt{2}H^-})\;.
\end{array}\label{Flat-3}
\end{equation*}  
These four coordinate currents $J_a$ get modified when compared to the flat case and this corresponds to the change for a nonstandard 4-geometry on $\mathbb{R}^4$.  

To see this let us turn to the realization of the ${\cal N}=4, \hat{c}=4$ algebra in terms of the $SU(2)_k\times U(1)_Q$ bosonic currents $J_a,a=1,2,3$, and their superpartners which are free fermionic fields $\Psi^a$. The result reads:
\begin{equation}
\begin{array}{c}
T=-\frac{1}{2}[\frac{2}{k+2}J_i^2+J^2_4-\Psi_a \partial \psi_a+Q\partial J_4]\\[5pt]
G_4= \sqrt{\frac{2}{k+2}}(J_i\Psi_i+\frac{1}{3}\epsilon_{ijl}\Psi_i\Psi_j\Psi_l) +J_4\Psi_4+Q\partial \Psi_4 \\[5pt]
G_i= \sqrt{\frac{2}{k+2}}(-J_i\Psi_4+\epsilon_{ijl}J_j\Psi_l+\epsilon_{ijl}\Psi_4\Psi_j\Psi_l) +J_4\Psi_i+Q\partial \Psi_i,i=1,2,3 \\[5pt]
S_i=\frac{1}{2}(\Psi_4\Psi_i+\frac{1}{2}\epsilon_{ijl}\Psi_j\Psi_l)\;.
\end{array}\label{SU-Con-Alg}
\end{equation}

Next we again complexify the generators and bosonise the free fermions by the scalar fields $H^+,H^-$:
\begin{equation}
\begin{array}{c}
T=-\frac{1}{2}[(\partial H^+)^2+(\partial H^-)^2+Q^2(J_1^2+J_2^2+J_3^2) +J_4^2+Q\partial J_4] \\[5pt]
G=\frac{G_1+iG_2}{\sqrt{2}}=-(\Pi_k^{\dag}e^{-\frac{i}{\sqrt{2}}H^-}+P_k^{\dag}e^{\frac{i}{\sqrt{2}}H^-})e^{\frac{i}{\sqrt{2}}H^+} \\[5pt]
\tilde{G}=\frac{G_4+iG_3}{\sqrt{2}}=(\Pi_k^+e^{\frac{i}{\sqrt{2}}H^-}-P_ke^{-\frac{i}{\sqrt{2}}H^-})e^{\frac{i}{\sqrt{2}}H^+} \\[5pt]
S_3=\frac{1}{\sqrt{2}}\partial H^+\:,\;S_{\pm}=e^{\pm i\sqrt{2}H^+}
\end{array}\label{SU-Con-Alg}
\end{equation}
where the coordinate currents now read:
\begin{equation}
\begin{array}{c}
P_k=Q(J_1+iJ_2) \\[5pt]
P_k^{\dag}=Q(-J_1+iJ_2) \\[5pt]
\Pi_k=J_4+iQ(J_3+\sqrt{2}\partial H^-) \\[5pt]
\Pi_k^{\dag}=-J_4+iQ(J_3+\sqrt{2}\partial H^-)\;.
\end{array}\label{SU-Con-Alg}
\end{equation}
Thus the modification of the coordinate currents when shifted from flat to exotic curved 4-d background can be described as: 
\begin{equation}
\begin{array}{c}
\partial \Phi_4 \to \partial x_4=J_4 \\[5pt]
\partial \Phi_1 \to J_1\\[5pt]
\partial \Phi_2 \to J_2 \\[5pt]
\partial \Phi_3 \to J_3+\sqrt{2}\partial H^-\;,
\end{array}\label{SU-Con-Alg}
\end{equation}
where as before $\Phi_a, a=1,2,3,4$ are flat bosonic currents and $J_a,a=2,3,4$ the currents of $SU(2)_k$ while $J_4$ is the current of the non-compact direction. In this way we have the $SU(2)_k\times \mathbb{R}$ background representing the ${\cal N}=4, \hat{c}=4$ algebra. The flat non-compact $a=1$ direction remains flat. This gives the proper algebraic structures underlying the change of the smoothness on $\mathbb{R}^4$.  The structures are localized on $S^3$ in the sense of Fig. \ref{fig-2}b and represent global effects of non-standard 4-smoothness on $\mathbb{R}^4$ when realized as a member of a given radial family. 

We will see that direct relation of the above realization with exact string theory backgrounds allows for the computations of some physical effects. These effects rely on the fact that exotic $\mathbb{R}^4$'s are non-flat Riemannian manifolds and some kind of gravity might be involved here. The algebraic approach allows for the calculation of the global quantum effects via superstring theory without knowing the exact (global) coordinates on $\mathbb{R}^4_k$ \cite{AsselmKrol2011f,Krol2010,Krol2011d}. 

\section{Superstring backgrounds realizations of algebraic ends of small $\mathbb{R}^4$}
Given the superconformal ${\cal N}=4,\hat{c}=4$ realization on $SU(2)_k\times \mathbb{R}_Q$ we turn naturally to superstring theory where the above linear dilaton is the part of exact 10-dimensional string background \cite{Antoniadis94,KK95,KK95a}. More is true, the flat 4-space realization of the algebra is also the part of the string background and the six-dimensional remaining CFT theory is the same in both cases, hence the change between 4 dimensional flat and non-flat parts are entirely grasped by the corresponding 10-d superstring calculations. Owing the algebraic ends $SU(2)_k\times \mathbb{R}$ assigned to exotic $\mathbb{R}^4_k$ from a given radial family and supposing that non-flat 4-geometries exist on these exotic Riemannian 4-manifolds such that the metrics may be gravitationally valid (see e.g. Ref. \cite{Sladkowski2001}), one relates the algebraic ends with superstring theory as the theory of quantum gravity. 
This strategy was indeed realized recently in the realm of closed heterotic superstring theory and some 'physical' results appeared \cite{AsselmKrol2011f,Asselmeyer-Maluga2010,Krol2011d,QG-2012}.

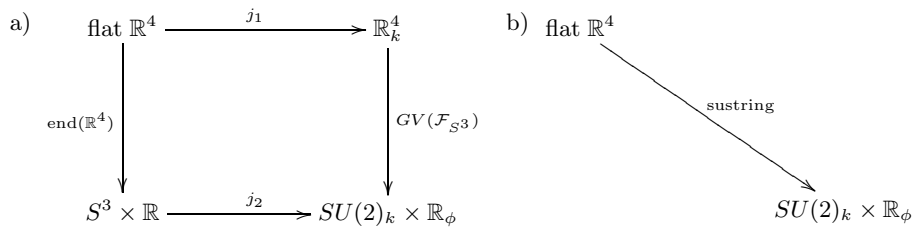
\begin{figure}[ht]
\centering
\begin{tabular}{cc}
a) \xymatrix{
   \rm{flat}\;\mathbb{R}^4
     \ar[rr]^{j_1}
     \ar[dd]_{\rm{end(\mathbb{R}^4)}} 
     && \mathbb{R}^4_k
     \ar[dd]^{GV({\cal{F}}_{S^3})} \\ \\
      S^3\times \mathbb{R}
     \ar[rr]^{j_2} 
      && SU(2)_k\times \mathbb{R}_{\phi} }  & \label{diag2}
b) \xymatrix{
\rm{flat}\; \mathbb{R}^4
\ar[ddrr]^{\rm{sustring}}
\\ \\ & & SU(2)_k\times \mathbb{R}_{\phi}}
\end{tabular}
\caption{For a given radial family: a) $j_1$ is the change of the standard smooth $\mathbb{R}^4$ to the exotic $\mathbb{R}^4_k$, $\rm{end(\mathbb{R}^4)}$ assigns the standard end to $\mathbb{R}^4$, $GV({\cal{F}}_{S^3})$ generates the $\rm{WZ}_k$-term from exotic $\mathbb{R}^4_k$ via $GV$ invariant of the codim.-1 foliation of $S^3$. b) The change of string backgrounds s.t. flat $\mathbb{R}^4$ part is replaced by the linear dilaton background $SU(2)_k\times \mathbb{R}_{\phi}$.}\label{fig-1}
\end{figure}

Thus, the general method applied in this task was to extend the relation between algebraic ends and SCFT's realized on them such that the 4-d parts of exact string theory backgrounds match precisely the algebraic ends. In that way gravity eventually described by Riemannian metrics on $\mathbb{R}^4_k$ from a given radial family, shares quantum perspective from superstring theory. We can illustrate this correspondence by the diagram as in Fig. \ref{fig-1}. 

Correct realization of the ${\cal N}=4,\hat{c}=4$ super CFT, suitable for the calculations, is again by the $\sigma$-model currents and fields on $SU(2)_k\times \mathbb{R}_{\phi}\times W^6$ where $W^6$ is the fixed 6-d part of the string backgrounds, which stays the same for flat and curved 4-d parts.

Let us, as the example and following Refs. \cite{AsselmKrol2011f,KK95}, briefly discuss the case of (almost) constant magnetic field switched on on 4-d part of the heterotic superstring background $\mathbb{R}^4 \times W^6$. In closed string theory even constant magnetic field causes the background to be curved and the correct choice is $SU(2)_k\times \mathbb{R}_{\phi}\times W^6$ \cite{KK95}. Thus, 4-d part of the backgrounds undergone the shift $\mathbb{R}^4\to SU(2)_k\times \mathbb{R}_{\phi}$ exactly as in Fig \ref{fig-1}b. This is correlated with the change of the smoothness structures from the standard to exotic $\mathbb{R}^4_k$ from a given radial family. 

The linear dilaton background realizes 4-d part $SU(2)_k\times \mathbb{R}_{\phi}$ as curved 4-manifold via $\sigma$-model and this SCFT allows for the calculation of the strong $H$-field effects on exotic $\mathbb{R}^4_k$ in the QG limit for even $k$ and for given radial family \cite{AsselmKrol2011f}. First, one switches to the CFT for even $k$ along the projection: $SU(2)_{k}\times\mathbb{R}_{Q}\to SO(3)_{k/2}\times \mathbb{R}_{Q}$ such that the  action for the heterotic $\sigma$-model reads:
\begin{equation}
S_{4}=\frac{k}{4}{\rm I}_{SO(3)}(\alpha,\beta,\gamma)+\frac{1}{2\pi}\int d^{2}z\left[\partial x^{0}\overline{\partial}x^{0}+\psi^{0}\partial\psi^{0}+\sum_{a=1}^{3}\psi^{a}\partial\psi^{a}\right]+\frac{Q}{4\pi}\int\sqrt{g}R^{(2)}x^{0}\label{eq:S4action}\end{equation}
where ${\rm I}_{SO(3)}(\alpha,\beta,\gamma)=\frac{1}{2\pi}\int d^{2}z\left[\partial\alpha\partial\overline{\alpha}+\partial\beta\partial\overline{\beta}+\partial\gamma\partial\overline{\gamma}+2cos\beta\partial\alpha\partial\gamma\right]$
in the Euler angles of $SU(2)=S^{3}$ and $Q$ is the charge of dilaton $\phi$.
Next, one includes the magnetic field $H$ and gravitational backreactions of it into the $\sigma$-model via suitable marginal deformations of CFT.  

Rewriting in terms of the currents, the marginal deformations take the general form of marginal vertex operators, i.e. $V_{m}=H(J^{3}+\psi^{1}\psi^{2})\overline{J}^{a}$ for magnetic $H$ part and $V_{gr}={\cal R}(J^{3}+\psi^{1}\psi^{2})\overline{J}^{3}$ for the gravitational ${\cal R}$ part with suitable choices made \cite{KK95,AsselmKrol2011f}. From such marginally deformed $\sigma$-model in the target $SU(2)_k\times \mathbb{R}_{\phi}$ one derives the new magneto-gravitational string theory background which depends now on the gravitational moduli $\lambda$. These moduli describe the geometry of squashed 3-spheres. On the other hand one starts with the semi-classical magneto-gravitational 3+1 dimensional configuration and arrive at the solutions - energy spectra, which, when compared with the string ones, give rise to the dictionary rules - one switches between string theory and semiclassical backgrounds and between the spectra \cite{KK95,AsselmKrol2011f}. In the regime of superstring theory, one has dependence on the level $k$ of the WZW model, which disappears in the classical point field theory limit. However, exotic geometries of $\mathbb{R}^4_{\frac{k}{2}}, k=0,2,4...$ generate dependence on $k$ via their algebraic ends. This is the source for shifting the border line between the regimes of quantum and classical gravities, since exotic $\mathbb{R}^4_k$ are 'classical' smooth curved Riemannian geometries and the dependence on $k$ is quantum string effect. Moreover, by this the non-trivial mass gap can appear in the theory due to exotic 4-smoothness, which is usually assigned to linear dilaton in the string regime \cite{AsselmKrol2011f,KK95}.

Let us turn to the modular data of the superstring theory backgrounds. The partition function of the $SO(3)$ WZW model at the level $k/2$ reads:
\begin{equation}
\Gamma(SO(3)_{k/2})=1/2\sum^1_{\gamma,\delta}Z_{so(3)}[^{\gamma}_{\delta}]
\end{equation}
where the following combination of characters of $SU(2)_k$ is used:
\begin{equation}
Z_{so(3)}[^{\alpha}_{\beta}]=e^{-i\pi \alpha\beta k/2}\sum^k_{l=0}e^{i\pi \beta l}\chi_l\overline{\chi}_{(1-2\alpha)l+\alpha k}\,.
\end{equation} 
The corresponding modular invariant partition functions $Z^W(\tau,\overline{\tau})$ for the $SU(2)_k\times \mathbb{R}_{\phi}\times W^6$, $k$ even, heterotic background can be obtained in the following form:
\begin{equation}\label{Z-W}
Z^W(\tau,\overline{\tau})={\rm Im}\tau^{3/2}|\eta|^6\frac{\Gamma(SO(3)_{k/2})}{V}Z_0(\tau,\overline{\tau})
\end{equation}   
where $Z_0(\tau,\overline{\tau})$ is the flat partition function of $\mathbb{R}^4\times W^6$ and $V=\frac{(k+2)^{3/2}}{8\pi}$ the quantum volume of $S^3$.

Thus, for a given radial family of exotic $\mathbb{R}^4$'s one assigns the modular expressions like above to the algebraic ends of $\mathbb{R}^4_{\frac{k}{2}}$ which are interpreted as parts of the superstring backgrounds. All the results are certainly achieved under the presence of 10-d supersymmetry. 
That is why the interesting problem is whether superconformal tools and modular characters in particular, allowing for deriving the physical valid results, are essential, or rather accidental, for our understanding of exotic $\mathbb{R}^4$. In the last section we try to approach this issue by the increasing the number of supersymmetries and giving up strict modularity thus, giving the insight into the relation of the algebraic ends and the actual topological ends of exotic $\mathbb{R}^4$ in terms of superconformal and quasi-modular tools. Noncommutative geometry deforms the algebras of modular forms what  further improves grasping exotic 4-spaces by these tools. 

\section{Representations of SCA ${\cal N}=4$, modular forms and exotic ends $S^3\times_{\Theta} \mathbb{R}$}
Let us, following Refs. \cite{Hikami2005,Hikami2005a,EguchiHikami2009}, recall the construction and some properties of the weight-$3/2$ modular form in terms of $SU(2)$ affine characters. This will give us the algebraic relation between $SU(2)_k$ and Brieskorn homology 3-spheres.  

First, given $2P$ theta functions parameterized by $a$ mod $2P$
\begin{equation}\label{tau-1}
\vartheta_{P,a}(z;\tau)=\sum_{n\in \mathbb{Z}}q^{\frac{(2Pn+a)^2}{4P}}e^{2\pi iz(2Pn+a)}\, ,
\end{equation}
where $q=e^{2\pi i \tau}, \tau \in \mathbb{H}$, 
they span $2P$-dimensional space of functions $f(z)$ with the property:
\begin{equation*}
\begin{array}{c}
f(z+1)=f(z)\\[5pt]
f(z+\tau)=e^{-4\pi iPz-2\pi i P\tau} f(z)\;.
\end{array}\label{tau-2}
\end{equation*}
The characters of the $SU(2)_k$ affine algebra can be written as:
\begin{equation}\label{tau-3}
\chi_{k,l}(z;\tau)=\frac{\vartheta_{k+2,2l+1}-\vartheta_{k+2,-2l-1}}{\vartheta_{2,1}-\vartheta_{2,-1}}(z,\tau)\, .
\end{equation}
Next, one defines the weight-$3/2$ modular form in terms of this character:
\begin{equation}\label{tau-4}
\Psi^{(a)}_P(\tau)=\chi_{P-2,\frac{a-1}{2}}(0;\tau)[\eta(\tau)]^3=\frac{\vartheta_{P,a}-\vartheta_{P,-a}}{\vartheta_{2,1}-\vartheta_{2,-1}}(0,\tau)[\eta(\tau)]^3\, , a\in \mathbb{Z}, 0< a < P \, ,
\end{equation}
here $\eta(\tau)$ is the Dedekind eta function on the upper half plane $\eta(\tau)=q^{\frac{1}{24}}\prod_{n=1}^{\infty}(1-q^n)$. The weight $\frac{3}{2}$ follows from the S-transformations of eta and theta functions: 
\begin{equation*}
\begin{array}{c}
\eta(\tau)=\sqrt{\frac{i}{\tau}}\eta(-\tau^{-1})\\[5pt]
\vartheta_{p,a}(z;\tau)=\sqrt{\frac{i}{\tau}}\frac{1}{\sqrt{2P}}e^{-\pi i\frac{2P}{\tau}z^2}\sum^{2P-1}_{b=0}e^{\frac{ab}{P}\pi i}\vartheta_{P,b}(\frac{z}{\tau};-\frac{1}{\tau})\;.
\end{array}\label{tau-5}
\end{equation*}
This last gives
\begin{equation*}\label{tau-6}
(\vartheta_{P,a}-\vartheta_{P,-a})(z;\tau)=i\sqrt{\frac{i}{\tau}}e^{-2\pi i\frac{P}{\tau}z^2}\sum^{P-1}_{b=1}S(P)_{ab}e^{\frac{ab}{P}\pi i}(\vartheta_{P,b}-\vartheta_{P,-b})(\frac{z}{\tau};-\frac{1}{\tau})
\end{equation*}
where the matrix $S(P)_{ab}=\sqrt{2/P}\sin{\frac{ab\pi}{P}}$.

Let us now turn to ${\cal N}=4,\hat{c}=6k$ superconformal algebra at general level $k$. This SCA is generated, similarly to our previous case of ${\cal N}=4, \hat{c}=4$, by the energy-momentum tensor, 4 supercurrents, and a triplet of currents which constitute the affine Lie algebra $SU(2)_k$. The unitary highest weight state $|\Omega\rangle$ is labelled by the conformal weight $h$ and the isospin $\ell$ fulfilling $0 \leq \ell \leq k/2$ with $\ell\in \mathbb{Z}/2$ and:
\begin{equation*}
\begin{array}{c}
L_0|\Omega\rangle = h|\Omega\rangle   \\[5pt]
T_0^3|\Omega\rangle=l|\Omega\rangle\;.
\end{array}\label{ch-1}
\end{equation*}
Given the representation $(k,h,\ell)$ on a Hilbert space ${\cal H}$ one calculates its character, as ${\rm ch}_{k,h,\ell}(z;\tau)={\rm Tr}_{{\cal H}}\l(e^{4\pi izT^3_0}q^{L_0-\frac{c}{24}} \l) $. There are two types of representations for ${\cal N}=4$ case, namely massless and massive. The first are also called BPS representations while massive - non-BPS. In the Ramond sector the characters of the massive representations, i.e. for $h>\frac{k}{4}, \ell =\frac{1}{2},1,...,\frac{k}{2}$, read:
\begin{equation}
{\rm ch}^R_{k,h,\ell}(z;\tau)=q^{h-\frac{\ell^2}{k+1}-\frac{k}{4}}\frac{[\theta_{10}(z;\tau)]^2}{[\eta(\tau)]^3}\chi_{k-1,l-\frac{1}{2}}(z;\tau).
\end{equation}
While massless characters, i.e. $h=\frac{k}{4}, \ell =0,\frac{1}{2},...,\frac{k}{2}$, again in the Ramond sector, one finds as: 
\begin{equation}
{\rm ch}^R_{k,\frac{k}{4},\ell}(z;\tau)=\frac{i}{\theta_{11}(2z;\tau)}\frac{[\theta_{10}(z;\tau)]^2}{[\eta(\tau)]^3}\sum_{\epsilon=\pm 1}\sum_{m\in \mathbb{Z}}\epsilon\frac{e^{4\pi i\epsilon((k+1)m+\ell)z}}{(1+e^{-2\pi i\epsilon z}q^{-m})}   q^{(k+1)m^2+2\ell m}.
\end{equation}
Here, as usual, $\theta_{11}(z;\tau)=\sum_{n\in\mathbb{Z}}q^{\frac{1}{2}(n+\frac{1}{2})^2}e^{2\pi i(n+\frac{1}{2})(z+\frac{1}{2})}$, $\theta_{10}(z;\tau)=\sum_{n\in\mathbb{Z}}q^{\frac{1}{2}(n+\frac{1}{2})^2}e^{2\pi i(n+\frac{1}{2})z}$ are the Jacobi theta functions and it holds: $(\vartheta_{2,1}-\vartheta_{2,-1})=-i\theta_{11}(2z;\tau)$.
Thus massless character at level-$k$ with isospin-0 in the $\widetilde{R}$-sector reads as:
\begin{equation}\label{ch-3}
{\rm ch}^{\widetilde{R}}_{k,\frac{k}{4},0}(z;\tau)=\frac{[\theta_{11}(z;\tau)]^2}{[\eta]^3}\frac{i}{\theta_{11}(2z;\tau)}\sum_{n\in \mathbb{Z}}\frac{1+e^{2\pi iz}q^n}{(1-e^{2\pi iz}q^{n})} .
\end{equation}
Now the point is to recognize the modular transformation properties of the massless characters. The problem with the massless characters as in (\ref{ch-3}) is that they have rather quasi-modular properties than strictly modular: these expressions have the form of so called Lerch sum. This last is connected with famous mock theta functions as first appeared in some of Ramanujan letters. Mock theta functions are not modular, these are quasi-modular in a precise sense which was not easy to uncover. Namely, the PHD thesis of Zweger \cite{Zwegers2002PHD} explains this subtle and difficult point, and states that: \emph{the mock theta function is a holomorphic part of the harmonic Maass form with weight 1/2}. The Maass form is a Jacobi form and can be presented as an eigenfunction of some second order differential operator. The understanding of these relations was in fact the true breakthrough in the field and gave strong tools applicable also in our case. In fact this was performed in a series of papers and in Ref. \cite{EguchiHikami2009} it was shown how to assign a Jacobi (Maass-type) form to each massless representation, such that these form fulfil hyperbolic second order Poisson differential equations. 

One can rephrase the result as: \emph{the massless characters (\ref{ch-3}) are the mock theta functions whose {\rm{shadow}} is (\ref{tau-4}) for $P=k$}. We try to explain briefly this point. 

Even though the massless characters (\ref{ch-3}) do not show modular properties, one can built their completions ${\widehat{\rm ch}}^{\widetilde{R}}_{k,\frac{k}{4},0}(z;\tau)$ which already do \cite{EguchiHikami2009}:
\begin{equation}\label{ch-4}
{\widehat{\rm ch}}^{\widetilde{R}}_{k,\frac{k}{4},0}(z;\tau)={\rm ch}^{\widetilde{R}}_{k,\frac{k}{4},0}(z;\tau)-\frac{1}{i\sqrt{2(k+1)}}\sum_{a=1}^kR^{(a)}_k(\tau )B_k^{(a)}(z;\tau)
\end{equation}
where $B^{(a)}_k(z;\tau)=\frac{[\theta_{11}(z;\tau)]^2}{[\eta(\tau)]^3}\frac{\vartheta_{P,a}-\vartheta_{P,-a}}{\vartheta_{2,1}-\vartheta_{2,-1}}(z;\tau)$ and $R^{(a)}_k(\tau)$ can be presented as a non-holomorphic 'period' integral, i.e.:
\begin{equation}\label{ch-5}
R^{(a)}_k(\tau)=\int_{-\overline{\tau}}^{i\infty}\frac{\Psi^{(a)}_k}{\sqrt{\frac{z+\tau}{i}}}dz .
\end{equation}
Here $\Psi^{(a)}_k$ is our vector-valued modular form of weight $3/2$ as appeared in (\ref{tau-4}). 
In the original case of mock-theta functions Zwegers observed that they allow modular completion, too and according to Zagier \cite{Zagier2006} one says that the massless character (\ref{tau-4}) are mock theta functions whose shadows are $\Psi^{(a)}_k$. 

On the other hand quasi-modular behavior of the massless characters of the SCA is the key toward calculating the invariants, like Witten-Reshetikhin-Turaev (WRT) or Chern-Simons (CS), of some 3-dimensional manifolds. We will show, following Refs. \cite{LawrZag1999,EguchiHikami2009,Hikami2005a,Hikami2005}, how to obtain WRT or CS invariants of the Brieskorn homology 3-spheres $\Sigma(p_1,p_2,p_3)$ from the massless characters of SCA. In fact these invariants are characterized by mock theta functions which indicates that indeed SCA ${\cal N}=4,2$ are essential here. The lower amount of supersymmetry in the algebra forbid the appearance of quasi-modularity as in mock theta functions. This is also the reason that exotic $\mathbb{R}^4$ is involved in this formalism. So, suitable amount of supersymmetry and conformal invariance are essential here.   

The weight-$3/2$ modular form in (\ref{tau-4}) one rewrites via its Fourier transformation as \cite{EguchiHikami2009}:
\begin{equation}\label{ch-6}
\Psi^{(a)}_P(\tau)=\frac{1}{2}\sum_{n\in \mathbb{Z}}n\psi^{(a)}_{2P}(n)q^{\frac{n^2}{4P}} 
\end{equation}
where 
\begin{equation}\label{ch-7}
\psi^{(a)}_{2P}(n)=\begin{cases}\pm 1, n=\pm a\, {\rm mod}\, 2P, & \\ 0, \rm{otherwise}.
\end{cases} 
\end{equation}
Let us redefine this as odd periodic function with modulus $2P$ in terms of triples $l_1,l_2,l_3$ of positive integers which are pairwise relatively prime and $1\leq l_j\leq p_j-1$ where we set $P=p_1p_2p_3$ and $\epsilon_j=\pm 1$:
 \begin{equation}\label{ch-8}
\psi^{(l_1,l_2,l_3)}_{2P}(n)=\begin{cases} 1, n=P\l(1+\sum_{j=1}^3\epsilon_j \frac{l_j}{p_j}\l) \, {\rm mod}\, 2P, \, {\rm where}\, \epsilon_1\epsilon_2\epsilon_3 = -1 & \\ -1, 
n=P\l(1+\sum_{j=1}^3\epsilon_j \frac{l_j}{p_j}\l)\, {\rm mod}\, 2P, \, {\rm where}\, \epsilon_1\epsilon_2\epsilon_3 = 1 & \\
0,\,\rm{otherwise}.
\end{cases} 
\end{equation}
Then, the Fourier transformation of this function reads:
\begin{equation}\label{ch-9}
\Psi^{(l_1,l_2,l_3)}_{\overline{p}}(\tau)=\frac{1}{2}\sum_{n\in \mathbb{Z}}n\psi^{(l_1,l_2,l_3)}_{2P}(n)q^{\frac{n^2}{4P}}. 
\end{equation}
Comparing with (\ref{tau-4}) we arrive at
\begin{lemma}(Prop. 2, Ref. \cite{Hikami2005})
The function $\Psi^{(l_1,l_2,l_3)}_{\overline{p}}(\tau)$ is a modular form of weight 3/2.  
\end{lemma}
The crucial now is to build the Eichler integral of $\Psi^{(l_1,l_2,l_3)}_{\overline{p}}(\tau)$ which is:
\begin{equation} 
 \widetilde{\Psi}_{\overline{p}}^{(l_1,l_2,l_3)}(\tau)=\sum_{n=0}^{\infty}\psi^{(l_1,l_2,l_3)}_{2P}(n)q^{\frac{n^2}{4P}}
\end{equation}
As shown in Ref. \cite{LawrZag1999} (see also Ref. \cite{Hikami2005}) the Eichler integral as above for $\tau \in \mathbb{Q}$ is not modular any longer but rather it is quasi-modular and has suitable asymptotic expansion at $\tau \to \frac{1}{N}$. Namely, defining, in analogy with (\ref{ch-5}), for $z\in \mathbb{H}^-$, the expression (period integral):
\begin{equation}\label{ch-10}
\widehat{\Psi}_{\overline{p}}^{(l_1,l_2,l_3)}(z)=\frac{1}{\sqrt{2Pi}}\int_{\overline{z}}^{\infty}\frac{\Psi^{(l_1,l_2,l_3)}_P(\tau)}{\sqrt{\tau-z}}d\tau
\end{equation}
and performing the $S$-transformation $z\to -\frac{1}{z}$ by suitable modular $\mathbf{S}$-matrix, we have:
\begin{equation}\label{ch-11}
\frac{1}{\sqrt{iz}}\sum_{(l'_1,l'_2,l'_3)}\mathbf{S}^{l'_1,l'_2,l'_3}_{l_1,l_2,l_3}\widehat{\Psi}_{\overline{p}}^{(l'_1,l'_2,l'_3)}{\l(}-\frac{1}{z}{\l)}+\widehat{\Psi}_{\overline{p}}^{(l_1,l_2,l_3)}(z)=\frac{1}{\sqrt{2Pi}}\int_{\alpha\searrow 0}^{\infty}\frac{{\Psi}_{\overline{p}}^{(l_1,l_2,l_3)}(\tau)}{\sqrt{\tau-z}}d\tau
\end{equation}
where $\alpha \in \mathbb{Q},z\in \mathbb{H}^-$. Now substituting (\ref{ch-9}) into (\ref{ch-10}) in the limit $z\to \alpha$, one has \cite{LawrZag1999,Hikami2005}: 
\begin{equation}\label{ch-12}
\widehat{\Psi}_{\overline{p}}^{(l_1,l_2,l_3)}(\alpha)=\widetilde{\Psi}_{\overline{p}}^{(l_1,l_2,l_3)}(\alpha).
\end{equation}
Thus, quasi-modularity of $\widetilde{\Psi}_{\overline{p}}^{(l_1,l_2,l_3)}(\alpha)$ in particular points $\alpha=\frac{1}{N}$ is inherited from this of $\widehat{\Psi}_{\overline{p}}^{(l_1,l_2,l_3)}(\frac{1}{N})$. As the consequence we have \cite{LawrZag1999,Hikami2005}:
\begin{theorem}(Theorem 9, Ref. \cite{Hikami2005}) \label{WRT-1}
Given the family of Brieskorn homology 3-spheres $\Sigma(p_1,p_2,p_3)$ each with the property that $\frac{p_1p_2+p_2p_3+p_1p_3}{P}< 1$, their WRT invariants, ${\rm{WRT}}_N(\Sigma((p_1,p_2,p_3)))$, are determined by:

\[e^{\frac{\pi i}{2N}(1-\frac{1}{P}+12(s_{231}+s_{132}+s_{123}))}\l( e^{\frac{2\pi i}{N}}-1\l){\rm{WRT}}_N(\Sigma((p_1,p_2,p_3)))=\frac{1}{2}\widetilde{\Psi}^{(1,1,1)}_{\overline{p}}(1/N).\]
\end{theorem} 
Here $s_{123}=s(p_1p_2,p_3)$ is the Dedekind sum $D(1,p_1p_2;p_3)$ and the Rademacher law holds: 
\[ D(1,b;c)+D(b,c;1)+D(1,c;b)=\frac{1}{12}\frac{1+b^2+c^2}{bc}-\frac{1}{4}.
\]
In the special case of the Poincar\'e 3-sphere ($\overline{p}=(2,3,5)$) the WRT invariant appears as \cite{LawrZag1999,Hikami2005}:
\begin{equation}
e^{\frac{2\pi i}{N}}\l(e^{\frac{2\pi i}{N}}-1\l){\rm{WRT}}_N(\Sigma(2,3,5))=1+\frac{1}{2}e^{-\frac{1}{60N}\pi i}\widetilde{\Psi}^{1,1,1}_{\overline{p}}(1/N).
\end{equation}
Given the Eichler integral $\widetilde{\Psi}^{(l_1,l_2,l_3)}_{\overline{p}}(z)$ one derives yet another invariant of the Brieskorn homology spheres, namely Chern-Simons (CS) invariant of $\Sigma(p_1,p_2,p_3)$:
\begin{theorem}(Theorem 7, Ref. \cite{Hikami2005})\label{CS-1}
In the limit value $N\in \mathbb{Z}$ of the Eichler integral it holds:
\[ \widetilde{\Psi}^{(l_1,l_2,l_3)}_{\overline{p}}(N)=-2e^{-2\pi i {\rm{CS}}(\Sigma(p_1,p_2,p_3))}N.\] 
\end{theorem}  
We see that the WRT invariants are derivable from the quasi-modularity of the massless characters of the SCA which again indicates on the connection with mock theta functions. This is strong indication that SCA ${\cal N}=4$, via its quasi-modular characters, can be related with exotic smoothness in dimension 4. 
\begin{theorem}\label{Modular-1}
From a family $(\theta_k),k=1,2,3...$ of Brieskorn homology 3-spheres one determines a family of smooth fake $(S^3\times_{\theta_k} \mathbb{R})$. Then, the invariants of these homology 3-spheres from the family as above, are calculable via the characters of SCA, ${\cal N}=4$, $\hat{c}=6r$, $r\in \mathbb{Z}$.
\end{theorem}

This follows from the results of Ref. \cite{Fre:79} where the correspondence of fake $S^3\times_{\Theta}\mathbb{R}$ with the homology 3-sphere $\Sigma_{\Theta}$ was established, and from the way how the WRT and CS invariants were obtained just before the statement of this theorem. 

This theorem can be illustrated as in Fig. \ref{fig-4} below:
\begin{figure}[ht]
\centering
\begin{tabular}{cc}
 \xymatrix{
 {\rm fake\, smooth\, ends:}\, & S^3 \times_{\theta_k} \mathbb{R} \ar[r] 
      \ar@{-->}[dd]^{\theta_l}
& S^3 \times_{\theta_l} \mathbb{R} \ar@{-->}[dd]^{\theta_k} \ar[r] & S^3 \times_{\theta_p} \mathbb{R} \ar@{-->}[dd]^{\theta_p} \, ... 
     \\ {\rm SCA}, {\cal N}=4,2,\hat{c}=6r \ar[rd]_{{\rm ch}^{\widetilde{R}}(z;\tau)}    
     \\ & {\rm WRT}(\theta_k) \ar[r] & {\rm WRT}(\theta_l) \ar[r] & {\rm WRT}(\theta_p)\, ...} 
    
\end{tabular}     
\caption{The homology 3-sphere $\theta_k$ determines fake smooth $S^3\times_{\theta_k}\mathbb{R}$ however, the WRT invariant of $\theta_k$ is calculated via massless characters of SCA, ${\cal N}=4$.} \label{fig-4}
\end{figure}

In what follows we would like to find a link between algebraic end and topological end of some exotic $\mathbb{R}^4_k$. The exoticness of the end of exotic $\mathbb{R}^4$ is rather of different kind than the fake $S^3\times_{\Theta}\mathbb{R}$. This last happens to be stable under the product by $\mathbb{R}$ \cite{Gompf84} while the exoticness on $\mathbb{R}^4$ is not, i.e. $\mathbb{R}^4_k\times \mathbb{R}$ is the unique standard smooth $\mathbb{R}^5$. Also, the periodic end smooth structure on $\mathbb{R}^4$ is not allowed for the members of the classes of compactly equivalent  $\mathbb{R}^4$'s \cite{Gom:89}.  

One can measure the complexity of any exotic 4-space $R^4$ via the minimum of first Betti numbers of the 3-dimensional closed manifolds $\Sigma \subset R^4$, which have the property that they \emph{separate from the infinity} any compact, co-dimension-0, submanifold $K\subset R^4$.  Such minimum value $e(R^4)$ over 1-st Betti numbers of $\Sigma$ for any small exotic $R^4$ has value $e(R^4)\leq 1$ \cite{Ganz2000}. The homology 3-spheres have certainly 0 as their first Betti numbers, so, in some cases they indeed could separate any compactum $K$ from the infinity, thus, indicating the complexity of the end. At present it is not known whether such embedded homology 3-spheres indeed exist in small exotic $\mathbb{R}^4$, though it is not excluded. In the case of large exotic $\mathbb{R}^4$ $e(R^4)> 1$ and such homology 3-spheres do not exist. Second, the homologies of the end $S^3\times \mathbb{R}$ cannot be represented by smoothly embedded $S^3$. Rather, certain homology 3-sphere is in order, but its fundamental group has to be non-trivially represented in $SO(3)$ \cite{DeMichFreedman1992}.

Still, any direct and invariant assignment of a single homology 3-sphere to exotic ends of small exotic $\mathbb{R}^4$'s is rather problematic. Possibly, one would assign a sequence of homology 3-spheres to the exotic ends to avoid end-periodicity.
Having the above in mind, let us conjecture at this stage that indeed exotic ends of some exotic $\mathbb{R}^4_k$ determine certain sequence of Brieskorn homology 3-spheres. Under this, rather rough supposition we can consider directly our previous SCA ${\cal N}=4$ massless characters calculations as giving the link between algebraic end and exotic end of some exotic $\mathbb{R}^4_k$ from the radial family. In the case one needs other, than homology 3-spheres, 3-manifolds, say with 1-st Betti number 1, their invariants still could be determined in principle from the SCA, ${\cal N}=4$ characters.
 
Following the discussion above, let us consider the family $(\mathbb{R}^4_l),l=l(k), k=1,2,3...$ of those $\mathbb{R}^4_l$ which have the fake end whose complexity determines sequences of homology Brieskorn 3-spheres, then:   
\begin{theorem}\label{Modular-2}
Given small exotic $\mathbb{R}^4_k$, $k=1,2,3,4,...$ from DeMichelis-Freedman family, one assigns algebraic ends $SU(2)_k\times \mathbb{R}$ to these. Suppose that for some indices $l\in \mathbb{N}$, exotic ends of $\mathbb{R}^4_l$ determine sequences of the homology 3-spheres $(\Sigma(l_1,l_2,l_3))_{l=l(k)}$. Then, the algebraic ends are related to these topological smooth ends via the SCA ${\cal N}=4, \hat{c}=6r$ algebra in a sense that one calculates the WRT invariants of $\Sigma(l_1,l_2,l_3)\in (\Sigma(l_1,l_2,l_3))_{l=l(k)}$ from the characters of the algebra, and this algebra is considered as the extension of the SCA $r=1$ $\hat{c}=4$ algebra which is realized on $SU(2)_k\times \mathbb{R}$.  
\end{theorem}

Let us fix a radial family. Then, the assignment of the algebraic ends to $\mathbb{R}^4_k$ from it was explained in Secs. \ref{RF} and \ref{2dCFT}. Under the supposition, we have the sequences of the Brieskorn homology 3-spheres assigned to some exotic $\mathbb{R}^4_{l(k)}$. Then, the determination of the WRT invariants of these spheres from SCA ${\cal N}=4, \hat{c}=6r$ was presented before the theorem. The SCA $r=1$ with $\hat{c}=4$ is indeed realized on $SU(2)_k\times \mathbb{R}$ as explained in Sec. \ref{4SCFT}. 

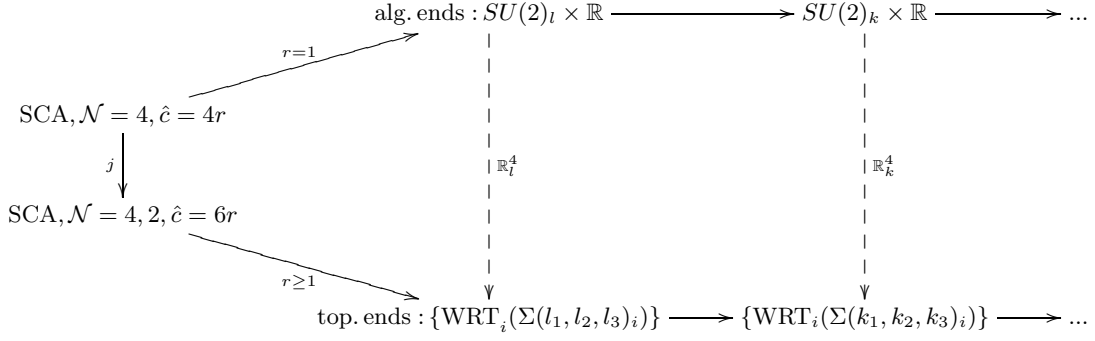
\begin{figure}[ht]
\centering
\begin{tabular}{cc}
 \xymatrix{
 & {\rm alg.\, ends:}\, SU(2)_l \times \mathbb{R} \ar[r] 
      \ar@{-->}[ddd]^{\mathbb{R}^4_l}
& SU(2)_k \times \mathbb{R} \ar@{-->}[ddd]^{\mathbb{R}^4_k} \ar[r] & ...  
     \\ {\rm SCA}, {\cal N}=4,\hat{c}=4r \ar[d]_{j} \ar[ru]^{r=1}
  \\ {\rm SCA}, {\cal N}=4,2,\hat{c}=6r  \ar[rd]_{r \geq 1} 
  \\ & {\rm top.\, ends:}\,{\{\rm WRT}_i(\Sigma(l_1,l_2,l_3)_i)\} \ar[r] & \{{\rm WRT}_i(\Sigma(k_1,k_2,k_3)_i)\} \ar[r] &  ...} 
    
\end{tabular}     
\caption{Exotic $\mathbb{R}^4_k$ from the fixed radial family - the relation between algebraic end and smooth (topological) end via superconformal algebras.} \label{fig-3}
\end{figure}

The relation between algebraic ends and topological ends via SCA is presented schematically at Fig. \ref{fig-3}. Still, more thorough understanding and description of exotic $\mathbb{R}^4_k$ via SCA characters requires constructions of the homology 3-spheres from the topological ends and development of the counterpart of elliptic genus for the smooth case of open manifolds. 

In the remaining part of the paper we are going to relate the algebraic end of exotic smooth $\mathbb{R}^4_k$ from the radial family of De Michelis-Freedman type with the topological end based on the cobordism classes of codimension-one foliations of a homology 3-sphere. 
Given the algebraic end $SU(2)_k\times \mathbb{R}$ of $\mathbb{R}^4_k$ one in fact has some codimension-one foliation of $S^3$ and this foliation is determined by the foliation of the boundary of the Akbulut cork. The foliations in question are the Thurston-type foliations of compact closed 3-manifolds and they depend on the surface of a polygon in the hyperbolic plane $\mathbb{H}^2$ and have non-vanishing Godbillon-Vey class \cite{AsselmeyerKrol2009,AsselmeyerKrol2011}. Namely, the following theorem holds:
\begin{theorem}(Theorem 4, Ref. \cite{AsselmeyerKrol2009})
Let $\Sigma$ be a compact 3-manifold without boundary. Every codimension-one
foliation ${\cal F}$ of the 3-sphere $S^3$ of Thurston type induces a codimension-one
foliation ${\cal F}_{\Sigma}$ of Thurston type on $\Sigma$. Every cobordism class $[{\cal F}]$ is represented by the Godbillon-Vey invariant $\Gamma_{\cal F}\in H^3(S^3;\mathbb{R})$. Moreover, let the cobordism class of a codimension-one foliation ${\cal F}_{\Sigma}$ of $\Sigma$ be $[{\cal F}_\Sigma ]$. Then, the GV number, $GV(\Sigma,{\cal F}_{\Sigma})\in H^3(\Sigma;\mathbb{R})$ is equal to the number $GV(S^3;{\cal F})\in \mathbb{R}$ and is thus, an invariant of $[{\cal F}_\Sigma ]$.
\end{theorem}
This theorem gives the direct assignment of the cobordism class $[{\cal F}_{\Sigma_k}]$ of the codimension-one foliation ${\cal F}_{\Sigma_k}$ on some homology 3-sphere $\Sigma_k$, to the algebraic end $SU(2)_k\times \mathbb{R}$ in a fixed radial family. The reason is the following: the cobordism class $[{\cal F}_{\Sigma_k}]$ involves as cobordant the foliation ${\cal F}_{\Sigma_k}$ and the codimension-one foliation of $S^3$, ${\cal F}_{S^3}$, and $S^3$ is chosen as the boundary of some 4-dimensional disk. Thus, $SU(2)_k$ is derivable precisely from the foliation of the boundary of the Akbulut cork which is the homology 3-sphere $\Sigma_k$ \cite{AsselmeyerKrol2009}. Thus, one can turn forth and back between the foliations of homology 3-sphere $\Sigma_k$ and $S^3$ within the cobordism class. The crucial task now would be to describe the foliations of $\Sigma_k$ by modular and superconformal tools.  

\begin{figure}[ht]
\centering
\begin{tabular}{cc}
 \xymatrix{
  {\rm alg.\, ends:}\, & SU(2)_l \times \mathbb{R} \ar@{-->}[r] 
      \ar@{-->}[dd]^{\mathbb{R}^4_l}
& SU(2)_k \times \mathbb{R} \ar@{-->}[dd]^{\mathbb{R}^4_k} \ar@{-->}[r] & SU(2)_p \times \mathbb{R} \ar@{-->}[dd]^{\mathbb{R}^4_p} \, ... 
     \\ {\rm SCA}, {\cal N}=4,\hat{c}=6r \ar[rd]_{\sigma -{\rm model\,\, on}\, {{\cal F}}_\Sigma} \ar[ru]^{S^3} \ar[rru] \ar[rrru]    
     \\ {\rm topol. ends:}\,& {{\cal F}}_{\Sigma_l} \ar@{-->}[r] & {{\cal F}}_{\Sigma_k} \ar@{-->}[r] & {{\cal F}}_{\Sigma_p}\, ...} 
    
\end{tabular}     
\caption{Algebraic ends $SU(2)_k\times \mathbb{R}$ correspond to a kind of non-linear $\sigma$-model on the foliation ${{\cal F}}_{\Sigma_k}$ of the homology 3-sphere $\Sigma_k$.} \label{fig-5}
\end{figure}
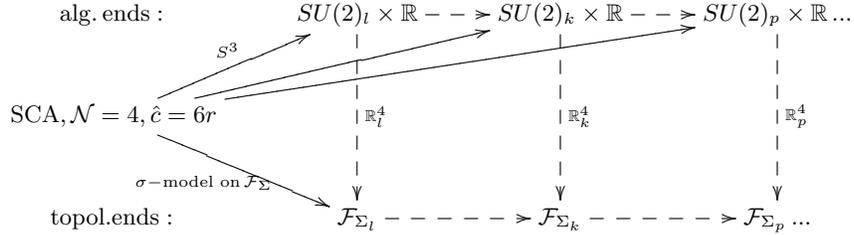

Again, the $S^3$ refers to algebraic end via $SU(2)_k$, while $\Sigma_k$ rather to the topological end of exotic $\mathbb{R}^4_k$ via the foliations. This is schematically depicted at Fig. \ref{fig-5}.  

Uncovering this relation can be performed by looking at the modification of the modular data, used before for characterizing homology 3-spheres, by the structure of the codimension-one foliations of the spheres. 
Thus, we enter the realm of noncommutative geometry where spaces of leaves of foliated manifolds are profound and basic examples of noncommutative spaces. Even though not all noncommutative algebras arise as the convolution algebras of some foliations, all three factors of $C^{\star}$-algebras do, and 'the most noncommutative' case, the factor III case, is realized by the foliations with non-vanishing Godbillon-Vey class. This is precisely the case at hand. Note that the factor I algebra is generated by the Reeb foliation whereas the factor II by the Kronecker foliations of torii and both have vanishing GV class. Thus, the modification of the modular data should refer to the noncommutative algebras generated by the spaces of leaves of codimension-one foliations of $\Sigma_k$ with non-vanishing GV class. The approach by Connes and Moscovici \cite{ConnesMosc2004,ConnesMosc2004a,ConnesMosc2005,ConnesMosc1998} explains this. To formulate the result let us fix the terminology more carefully. More details can be found in the \ref{app}.

The space $S_k(1)$ of cusp forms of level 1 and weight $k$ consists of all functions $f$ which are holomorphic on the upper half plane $H=\{z\in\mathbb{C}: {\rm Im}(z)>0  \}$ such that for all 
$\left(\begin{array}{cc}
a & b\\
c & d\end{array}\right)\in {\rm SL}(2,\mathbb{Z})$ it holds:
\begin{equation*}
f\left(\frac{a\tau+b}{c\tau+d}\right)=(c\tau+b)^kf(\tau)\,.
\end{equation*}
Let $\Gamma_1(N)$ be the subgroup of ${\rm SL}(2,\mathbb{Z})$ of matrices which are
equal $\left(\begin{array}{cc}
1 & \star \\
0 & 1\end{array}\right)$ with entries reduced modulo $N$, then the space of \emph{modular forms}, $M_k(\Gamma_1(N))$, of level $N$ and weight $k$ reads:
\begin{equation}\label{Mod-1}
M_k(\Gamma_1(N))=\{f:f(\gamma \tau)=(c\tau + d)^kf(\tau)\, {\rm for\, all\,} \gamma\in\Gamma_1(N)\}\,.
\end{equation}
where $f$ are holomorphic on $H$ and on all cusps. 
The space $S_k(\Gamma_1(N))\subset M_k(\Gamma_1(N))$ of \emph{cusp forms} of level $N$ and weight $k$ contains those modular forms which vanish at cusps. When one applies $\left(\begin{array}{cc}
1 & 1\\
0 & 1\end{array}\right)\in {\rm SL}(2,\mathbb{Z})$ in (\ref{Mod-1}) the result is $f(z)=f(z+1)$ for every modular form $f$. Then, the function $F$ of $q=e^{2\pi iz}$ such that $F(q)=f(z)$ can be determined. When $f(z)$ is holomorphic and vanishes at $\infty$, $F(z)$ can be extended to a holomorphic function such that $F(0)=0$ and $F$ is defined on $\{z:|z|<1 \}$. In this case $f$ is recovered from the Fourier expansion coefficients as: $f(q)=\sum_{n=1}^{\infty}a_nq^n$. This is essentially the representation of the modular functions we used at the first part of this section in determining WRT and CS invariants.

Modular forms $f$ of weight $k$ at the fixed level $N$ with respect to \emph{any} congruence subgroup $\Gamma(N)$ of ${\rm SL}(2,\mathbb{Z})$ form graded algebras of forms of level $N$:
\begin{equation*} 
{\cal M}(\Gamma(N))=\sum_{k \geq 1}^{\oplus}{\cal M}_k(\Gamma(N))\,.
\end{equation*} 
The important fact is that the ${\cal M}_k(\Gamma(N))$ is finite dimensional for every $k\in \mathbb{N}$ and $\Gamma(N)$ \cite{Zagier2008}.
The full algebra, ${\cal M}$, of modular forms of \emph{all levels} is then, created as the projective limit (see e.g. Ref. \cite{ConnesMosc2004}):
\begin{equation*}
{\cal M}=\lim_{\overrightarrow{{N \to \infty}}}{\cal M}(\Gamma(N))\,.
\end{equation*} 
This is precisely the modular algebra which encodes the modifications due to the nontrivial GV class of the codimension-one foliations of closed 3-manifolds like $\Sigma$. 
Namely, given the codimension-one foliation of $\Sigma$ with non-zero GV class, one generates the Hopf algebra ${\cal H}_1$ (see the \ref{app}) which interprets the GV class as its Hopf cyclic cocycle. We can formulate:  
\begin{theorem}
For DeMichelis-Freedman radial family of exotic $\mathbb{R}^4$'s the exotic smoothness structures of the members determine non-holomorphic canonical deformation of the modular functions from ${\cal M}$. This deformation is determined by the action of the Hopf algebra ${\cal H}_1$ on crossed product ${\cal M}\rtimes {\rm GL}^+(2,\mathbb{Q})$ where the GV class of  the codimension-one foliations of $\Sigma$ is interpreted as the cyclic cocycle of the ${\cal H}_1$.  
\end{theorem}

The relation of exotic $\mathbb{R}^4$'s from the radial family with codimension-one foliations of some homology 3-sphere with nonvanishing Godbillon-Vey class, was analyzed in Ref. \cite{AsselmKrol2011c} and is reformulated in Theorem \ref{thm:codim-1-foli-radial-fam} in this paper. Thus, we have to show that the modification of ${\cal M}$ caused by the foliation with non-vanishing Godbillon-Vey class, takes form precisely as in the theorem. But this is the construction by Connes and Moscovici \cite{ConnesMosc1998,ConnesMosc2005,ConnesMosc2004} and in particular the non-holomorphic action of the GV class on ${\cal M}$ is as in Th. \ref{GV}. We sketch some points of the Connes-Moscovici construction in \ref{app}.
 
The interesting further part would be the calculation of the semi-modular data directly from some superconformal field theory in the exotic target. Since the modular algebra is now twisted due to the noncommutative ingredient of the foliations, the correct setup would be a kind of \emph{(smooth) superconformal non-commutative field theory}. 

In the advent of such theory let us comment here on the important for that purpose 
attempt to formulate non-linear $\sigma$-model with a foliation as the target \cite{Zois2000,Zois2000b}. The foliation is the one of a compact closed 3-manifold such as $\Sigma$. This kind of $\sigma$-model contains the counterpart which corresponds to ordinary $\sigma$-model on a manifold target but this is modified due to the noncommutative geometry of the foliation. The proper pairing of the classes leads to the invariant of the $\sigma$-model in the noncommutative geometry of the foliation. Interestingly, such invariant is proposed as a way towards the M-theory lagrangian \cite{Zois2000,Zois2000b} which corresponds to the yet unknown M-field theory. 

In general SCFT's, ${\cal N}=4,2$ in 2 dimensions are usually considered as basic tools for constructing superstring vacua, they are also interesting subjects in their own
and have been studied extensively. The results in the paper show that they might be also important in exploring exotic $\mathbb{R}^4$. The elliptic genera of such SCFT's usually allow for exact computations. Possibly, a kind of \emph{smooth elliptic genera} should be considered here which extend the topological genera. However, a substantial further work on exotic $\mathbb{R}^4$ is required. Better understanding of geometry and dynamics of string theory as seen from dimension four and from modular perspective, should emerge. Also the way how the radial family of exotic $\mathbb{R}^4$'s is affecting the string dynamics would be important.

That is why our next step is to represent the GV-modified quasi-modular expressions as correlation functions in superstring theory. This is performed in the accompanying paper.
  
\section*{Acknowledgments}
JK thanks Sebastian Zaj\k{a}c for the discussion and help with improving the text.
\appendix

\section{Godbillon-Vey class and cyclic cohomologies of Hopf algebra and its action on modular forms}\label{app}

To any congruence subgroup $\Gamma$ of ${\rm SL}(2,\mathbb{Z})$ one associates ${\cal A}(\Gamma)$, the modular Hecke algebra of level $\Gamma$ \cite{ConnesMosc2004}. ${\cal A}(\Gamma)$ is a crossed product algebra and by this extends the commutative ring of Hecke operators and the algebra ${\cal M}(\Gamma)$ of $\Gamma$-modular forms.   
In terms of ${\cal M}$, the algebra of modular forms of arbitrary level, ${\cal A}(\Gamma)$ consists of maps with finite domain of equivalence classes: $F:\Gamma \backslash {\rm GL}^+(2,\mathbb{Q})\to {\cal M}$ such that for $\alpha \in {\rm GL}^+(2,\mathbb{Q})$ one has $\Gamma\alpha\mapsto F_{\alpha}$ and the covariance condition is satisfied:
\begin{equation}\label{-1}
F_{\alpha\gamma}(z)=F_{\alpha}\left(\frac{g_1z+g_2}{g_3z+g_4}\right)(g_3z+g_4)^{-1}=: F_{\alpha}|\gamma ,\; \forall \gamma = \left(\begin{array}{cc}
g_1 & g_2\\
g_3 & g_4\end{array} \right)
 \in \Gamma , \, \alpha \in {\rm GL}^+(2,\mathbb{Q}) \,. 
\end{equation} 

We want to explain briefly the Connes-Moscovici construction of the Hopf algebra ${\cal H}_1$ which represents the GV class of the codimension-1 foliations as Hopf cyclic class for ${\cal H}_1$. The next step is to show that and how ${\cal H}_1$ acts on the crossed product ${\cal M}\ltimes {\rm GL}^+(2,\mathbb{Q})$ and on ${\cal A}(\Gamma)$. 
This action shows how the modular algebra structure encodes the GV classes of the codimension-one foliations. 

The elements of the crossed product ${\cal M}\ltimes {\rm GL}^+(2,\mathbb{Q})$ can be represented by finite sums of monomials-symbols of the form:
\[\sum fU^{\star}_\gamma \, ,\;\; {\rm where}\; f\in {\cal M}\; {\rm and}\;  \gamma \in {\rm GL}^+(2,\mathbb{Q}) \,. \] Then the product is given by:
\[fU^{\star}_{\alpha} \cdot g U^{\star}_{\beta} =(f\cdot g|\alpha)U^{\star}_{\beta \alpha} \] and the $g|\alpha$ notation is defined in (\ref{-1}).

Given a codimension-one foliation of 3-d closed manifold $\Sigma$ one builds the Hopf algebra ${\cal H}_1$ such that it represents as its cyclic cocycles the GV class of the foliation and the transverse fundamental class thus, serving as the proper algebraic representation for the non-commutative geometry of the foliation. As the algebra ${\cal H}_1$ is the enveloping algebra of certain Lie algebra with the basis $\{X,Y,\delta_n;n\geq 1\}$ and the following commutators:
\begin{equation*}
[Y,X]=X,\,[Y,\delta_n]=n\delta_n,\, [X,\delta_n]=\delta_{n+1},\, [\delta_k,\delta_l]=o,\;\; n,k,l\geq1\,.
\end{equation*}  
The Hopf algebra structure is given by the coproduct $\Delta:{\cal H}_1\to {\cal H}_1\otimes {\cal H}_1$ which on generators is defined as:
\begin{equation*}
\Delta Y=Y\otimes 1 + 1\otimes Y,\; \Delta X=X\otimes 1+ 1\otimes X + \delta_1\otimes Y,\; \Delta \delta_1 =\delta_1\otimes 1 + 1\otimes \delta_1\,.
\end{equation*}
with the multiplicativity property:
\[ \Delta(h^1h^2)=\Delta (h^1)\cdot \Delta(h^2),\,h^1,h^2 \in {\cal H}_1; \]
the antipode $S$: 
\[S(Y)=-Y,\, S(X)=-X+\delta_1,\, S(\delta_1)=-\delta_1 \]
with the anti-isomorphism property:
\[S(h^1h^2)=S(h^2)S(h^1);\]
the counit $\epsilon(h)={\rm constant\; term}(h),\, h\in {\cal H}_1\,.$

The meaning of ${\cal H}_1$ relies on its natural action on crossed products as ${\cal A}(\Gamma)$ or ${\cal M}\ltimes {\rm GL}^+(2,\mathbb{Q})$ which shows that ${\cal H}_1$ serves as the symmetry of the transverse geometry of the foliations. Let us discuss this point \cite{ConnesMosc2004}.

Let us choose a discrete subgroup of the group of orientation preserving diffeomorphisms, $ \Gamma\subset {\rm Diff}^+(M^1)$, of a given 1-dimensional manifold $M^1$. Let $J^1_+(M^1)$ be the oriented 1-jet bundle over $M^1$. The coordinates on $J^1_+(M^1)$ are given by the Taylor expansion $j(s)=y+sy_1+...,\, y_1>0$ then, diffeomorphisms $\phi\in {\rm Diff}^+(M^1)$ act as: $\phi(y,y_1)=(\phi(y),\phi'(y)\cdot y_1)$. Then, ${\cal H}_1$ acts on the crossed product algebra ${\cal A}_{\Gamma}=C^{\infty}_c(J^1_+(M^1))\rtimes \Gamma$, as follows:
\[Y(fU^{\star}_{\phi})=y_1\frac{\partial f}{\partial y_1}U^{\star}_{\phi},\; X(fU^{\star}_{\phi})=y_1\frac{\partial f}{\partial y}U^{\star}_{\phi},\; \delta_n(fU^{\star}_{\phi})=y^n_1\frac{d^n}{dy^n}\left({\rm log}\frac{d\phi}{dy}\right)fU^{\star}_{\phi}\,. \]
There exists the volume form $\frac{dy\wedge dy_1}{y_1^2}$ on $J^1_+(M^1)$ which is ${\rm Diff}^+(M^1)$ invariant such that the following trace on ${\cal A}_{\Gamma}$ is defined: \begin{equation}\label{trace}
\tau(fU^{\star}_{\phi})= 
\begin{cases}
\int_{J^1_+(M^1)}f(y,y_1)\frac{dy\wedge dy_1}{y_1^2}\; {\rm if}\;\phi=1 \\
0\; {\rm if}\; \phi \neq 1\, ,
\end{cases}
\end{equation}
To define a cyclic cohomology $HC^{\star}({\cal H}_1)$ of ${\cal H}_1$ one needs first, a \emph{modular pair in involution} $(\nu , \sigma =1)$ \cite{ConnesMosc1998} where $\nu \in {\cal H}_1^{\star}$ is some modular character given by:
\[\nu (Y)=1,\, \nu (X)=0,\, \nu (\delta_n)=0\] and $\sigma = 1$ is a vector (group-like) in ${\cal H}_1$, and second, the $\nu$-invariant trace $\tau : {\cal A}_{\Gamma}\to \mathbb{C}$ with respect to the action ${\cal H}_1\otimes {\cal A}_{\Gamma}\to {\cal A}_{\Gamma}$. The invariant trace is precisely $\tau$ from (\ref{trace}) above, since it holds: \[ \tau(h(a))=\nu(h)\tau(a), \; h\in {\cal H}_1\,. \] 

When modifying $S$ to $\tilde{S}=\nu \star S$ one has $\tilde{S}^2={\rm Id}$ and such that:
\[\tilde{S}(\delta_1)=-\delta_1,\; \tilde{S}(Y)=-Y+1,\; \tilde{S}(X)=-X+\delta_1Y\,. \]

Let $\texttt{h}_1$ be the Lie algebra of formal vector fields on $\mathbb{R}^1$. Its Gelfand-Fuchs cohomology reads $H^{\star}(\texttt{h}_1,\mathbb{C})$. The cyclic cohomology and periodic cyclic cohomology of the Hopf algebra ${\cal H}_1$ are $HC^1({\cal H}_1)$ and $PHC^1({\cal H}_1)$, correspondingly. One of the results of Ref. \cite{ConnesMosc1998} is the isomorphism $\kappa^{\star}_1$ of the Gelfand-Fuchs and periodic cyclic cohomologies as above:
\[ \kappa^{\star}_1:H^{\star}(\texttt{h}_1,\mathbb{C})\to PHC^1({\cal H}_1) \,.  \]
Then it holds:
\begin{theorem}(Prop. 3, Ref. \cite{ConnesMosc2004})
The element $\delta_1\in {\cal H}_1$ is a Hopf cyclic cocycle such that it generates nontrivial class $[\delta_1]\in HC^1({\cal H}_1)$.

Under the isomorphism $\kappa ^{\star}_1$, $[\delta_1]$ is the image of the Godbillon-Vey class in $PHC^1({\cal H}_1)$ and is the generator for this periodic cohomologies.
\end{theorem}
Next one derives the important for this paper action of ${\cal H}_1$ on ${\cal M}\ltimes {\rm GL}^+(2,\mathbb{Q})$. The result reads:
\begin{theorem}(Prop. 7, Ref. \cite{ConnesMosc2004})\label{GV}
The unique Hopf action of ${\cal H}_1$ on ${\cal M}\ltimes {\rm GL}^+(2,\mathbb{Q})$ is determined by
\[X(fU^{\star}_{\gamma})=X(f)U^{\star}_{\gamma}, \; Y(fU^{\star}_{\gamma})=Y(f)U^{\star}_{\gamma},\; \delta_1(fU^{\star}_{\gamma})=\mu_{\gamma}\cdot fU^{\star}_{\gamma}\, ,  \] 
\end{theorem}
where the factor $\mu_{\gamma}$ is an Eisenstein series of weight 2 for every $\gamma = \left(\begin{array}{cc}
a & b\\
c & d\end{array} \right) \in {\rm GL}^+(2,\mathbb{Q})$ and can be written as:
\[\mu_{\gamma}(z)=\frac{1}{2\pi^2}\left(G_2^{\star}|\gamma(z)-G_2^{\star}(z)+\frac{2\pi ic}{cz+d}\right)   \]. The $G^{\star}_2=\frac{\pi^2}{3}-8\pi^2\sum_{m,n\leq 1}me^{2\pi imnz}$ is the holomorphic Esenstein series of weight 2 which fails to be modular, i.e.:
\[G^{\star}_2|\alpha (z)=G^{\star}_2(z)-\frac{2\pi ic}{cz+d},\; \alpha =  \left(\begin{array}{cc}
a & b\\
c & d\end{array} \right)\in \Gamma (1)\,. \]
In this way the GV class of the foliation is represented by the semi-modular object as above and by the modification of the structure of modular algebra ${\cal M}$. 

The corresponding action of $\delta_1$ from ${\cal H}_1$ on ${\cal A}(\Gamma)$ is derived from the general case (Theorem 12, Ref. \cite{ConnesMosc2004}):
\[\delta_n(F)_{\alpha}=X^{n-1}(\mu_{\alpha}) \cdot F_{\alpha} \,, \; \forall F\in {\cal A}(\Gamma),\;\alpha \in {\rm GL}^+(2,\mathbb{Q}) \]
as \[\delta_1(F)_{\alpha}=\mu_{\alpha} \cdot F_{\alpha}\,. \]




\begin{thebibliography}{10}

\bibitem{Asselmeyer-Maluga2010}
T.~Asselmeyer-Maluga,
  {\em Class. Q. Grav.} \textbf{27}, 165002, 2010,
   arXiv:1003.5506.

\bibitem{Asselm-Krol-2011}
T.~Asselmeyer-Maluga and J.~Kr\'ol, {T}he modification of the energy spectrum of charged particles by exotic open 4-smoothness via superstring theory,
  arXive:1112.4882.

\bibitem{Krol:04a}
J.~Kr{\'o}l,
  {\em Found. Phys.} \textbf{34}, 361, 2004.

\bibitem{Krol:04b}
J.~Kr{\'o}l,
  {\em Found. Phys.} \textbf{34}, 843, 2004.

\bibitem{Krol:2005}
J.~Kr{\'o}l, {M}odel theory and the {AdS/CFT} correspondence, presented at the IPM String School and Workshop, Queshm Island, Iran, 05-14. 01. 2005, arXiv:hep-th/0506003.

\bibitem{Wit:89}
E.~Witten, {\em Commun. Math. Phys.} \textbf{121}, 351, 1989.

\bibitem{AssKrol2010ICM}
T.~Asselmeyer-Maluga and J.~Kr{\'o}l, Small exotic smooth $R^4$ and string theory,
 in {\em International Congress of Mathematicians ICM, Hyderabad, India 2010, Short Communications Abstracts Book}, R. Bathia (Ed.), Hindustan Book Agency, p. 400 (2010).

\bibitem{AsselmKrol2011f}
T.~Asselmeyer-Maluga, P.~Gusin and J.~Kr\'ol, {T}he modification of the energy spectrum of charged particles by exotic open 4-smoothness via superstring theory,
  will appear in {\em Int. J. Geom. Meth. Mod. Phys.}
  \textbf{10} No 1, 2013, arXiv: 1109.1973.

\bibitem{Zwegers2002PHD}
S.~P. Zwegers, Ph.D. thesis, Universiteit Utrecht, 2002.

\bibitem{LawrZag1999}
R.~Lawrence and D.~Zagier,
{\em Asian J. Math.} \textbf{3}, 93, 1999.

\bibitem{Hikami2005}
K.~Hikami, {\em Int. J. Math.} \textbf{16}, 661 2005, arXiv:math-ph/0405028.

\bibitem{Hikami2005a}
K.~Hikami, {\em Regular and Chaotic Dynamics}
  \textbf{10}, 509, 2005 arXiv:math-ph/0506073.

\bibitem{Zagier2006}
D.~Zagier, {\em Seminaire Bourbaki} \textbf{986}, 2006.

\bibitem{AsselmeyerKrol2009}
T.~Asselmeyer-Maluga and J.~Kr\'ol, {A}belian Gerbes, generalized geometries and Exotic {$R^4$}, 2009, arXiv: 0904.1276.

\bibitem{Asselmeyer2007}
T.~Asselmeyer-Maluga,
C.~H.~Brans, {\em Exotic {S}moothness and {P}hysics}, World Scientific, Singapore, (2007).

\bibitem{Scorpan2005}
A.~Scorpan, {\em The {W}ild {W}orld of 4-{m}anifolds} AMS, USA, (2005).

\bibitem{DeMichFreedman1992}
S.~DeMichelis and M.~H.~Freedman, {\em J. Diff. Geom.} \textbf{35}, 219, 1992.

\bibitem{AsselmeyerKrol2009a}
T.~Asselmeyer-Maluga and J.~Kr{\'o}l, {G}erbes on Orbifolds and Exotic Smooth {$R^4$}, 2009, arXiv: 0911.0271.

\bibitem{AsselmeyerKrol2010}
T.~Asselmeyer-Maluga and J.~Kr{\'o}l, {E}xotic smooth $\mathbb{R}^4$, noncommutative algebras and quantization, 2010, arXiv: 1001.0882.

\bibitem{AsselmeyerKrol2011}
T.~Asselmeyer-Maluga and J.~Kr\'ol,
 {\em Int. J. Mod. Phys. A} \textbf{26}, 1375, 2011, arXiv:1101.3169.

\bibitem{AsselmeyerKrol2011b}
T.~Asselmeyer-Maluga and J.~Kr\'ol,
{\em Int. J. Mod. Phys. A} \textbf{26}, 3421, 2011, arXiv:1105.1557.

\bibitem{AsselmKrol2011c}
T.~Asselmeyer-Maluga and J.~Kr\'ol, {C}onstructing a quantum field theory from spacetime, 2011, arXiv:1107.3458.

\bibitem{AsselmKrol2011d}
T.~Asselmeyer-Maluga and J.~Kr\'ol,
{\em Int. J. Geom. Meth. Mod. Phys.} \textbf{9}, 2012, arXiv:1102.3274.

\bibitem{Krol2010}
J.~Kr{\'o}l, {\em Ann. Phys. (Berlin)} \textbf{19}, No. 3, 2010.

\bibitem{Krol2010b}
J.~Kr{\'o}l, {\em Acta. Phys. Pol. B} \textbf{40}(11), 3079, 2009.

\bibitem{Krol2011c}
J.~Kr{\'o}l, {\em Acta. Phys. Pol. B} \textbf{42}(11), 2335, 2011.

\bibitem{Krol2011d}
J.~Kr\'ol, {\em Acta. Phys. Pol. B} \textbf{42}(11), 2343 2011.

\bibitem{Fre:79}
M.~Freedman,
{\em Ann. of Math.} \textbf{110}, 177, 1979.

\bibitem{Antoniadis94}
I.~Antoniadis, S.~Ferrara and C.~Kounnas, {\em Nucl.Phys. B} \textbf{421}, 343, 1994,
 {C}ERN-TH.7148/94, arXiv:hep-th/9402073.

\bibitem{KK95}
E.~Kiritsis and C.~Kounnas, {\em Nucl. Phys. B} \textbf{456}, 699, 1995, arXiv:hep-th/9508078.

\bibitem{KK95a}
E.~Kiritsis and C.~Kounnas, {\em Nucl. Phys. B} \textbf{442}, 472, 1995, arXiv:hep-th/9501020v5.

\bibitem{Sladkowski2001}
J.~S{\l}adkowski, {\em Int. J. Mod. Phys. D} \textbf{10}, 311, 2001.

\bibitem{QG-2012}
J.~Kr\'ol, Quantum gravity insight from smooth 4-geometries on trivial $\mathbb{R}^4$, in \emph{Quantum Gravity}, Rodrigo
  Sobreiro (Ed.), ISBN: 978-953-51-0089-8, InTech, Available from:
  http://www.intechopen.com/articles/show/title/quantum-gravity-insights-from-smooth-4-geometries-on-trivial-r4, 2012.

\bibitem{EguchiHikami2009}
T.~Eguchi and K.~Hikami, {\em J.Phys.A} \textbf{42}, 304010, 2009, arXiv:0812.1151.

\bibitem{Gompf84}
R.~Gompf, {\em J. Diff. Geom.} \textbf{18}, 317, 1984.

\bibitem{Gom:89}
R.~Gompf, {\em Top. Appl.} \textbf{32}, 141, 1989.

\bibitem{Ganz2000}
S.~Ganzel,
{\em Topology Proceedings} \textbf{30}, 223, 2000.

\bibitem{ConnesMosc2004}
A.~Connes, and H.~Moscovici,
 {\em Moscow Math. J.} \textbf{4}, 2004.

\bibitem{ConnesMosc2004a}
A.~Connes, and H.~Moscovici,
{\em Moscow Math. J.} \textbf{4}, 111, 2004.

\bibitem{ConnesMosc2005}
A.~Connes, and H.~Moscovici, {T}ransgressions of the Godbillon-Vey class and Rademacher functions, 2005, math/0510683.

\bibitem{ConnesMosc1998}
A.~Connes, and H.~Moscovici,
 {\em Commun. Math. Phys.} \textbf{198}, 199, 1998.

\bibitem{Zagier2008}
D.~Zagier, Elliptic modular forms and their applications, in {\em The 1-2-3 of Modular Forms}, Springer, 2008. 

\bibitem{Zois2000}
I.~P.~Zois, {\em Commun.Math.Phys.} \textbf{209}, 757, 2000, arXiv:hep-th/9904001.

\bibitem{Zois2000b}
I.~P.~Zois, arXiv:hep-th/0006169.

\end{thebibliography}

\end{document}